\documentclass[a4paper,11pt]{article}
\pdfoutput=1 

\usepackage{jinstpub} 

\usepackage{lineno}
\usepackage{siunitx}
\usepackage{subfigure} 

\title{\boldmath Optimizing Charge Transport Simulation for Hybrid Pixel Detectors}

\author[1]{X.~Xie\note{Corresponding author.},} \author{R.~Barten,} \author{A.~Bergamaschi,} \author{B.~Braham,} \author{M.~Br\"uckner,} \author{M.~Carulla,} \author{R.~Dinapoli,} \author{S.~Ebner,} \author{K.~Ferjaoui,} \author{D.~Greiffenberg,} \author{S.~Hasanaj,} \author{J.~Heymes,} \author{V.~Hinger,} \author{T.~King,} \author{P.~Kozlowski,} \author{C.~Lopez-Cuenca,} \author{D.~Mezza,} \author{K.~Moustakas,} \author{A.~Mozzanica,} \author{K.A.~Paton,} \author{C.~Ruder,} \author{B.~Schmitt,} \author{P.~Sieberer,} \author{D.~Thattil,} \author{J.~Zhang,} \author{and~E.~Fr\"ojdh}

\affiliation{Paul Scherrer Institut,\\5232 Villigen PSI, Switzerland}

\emailAdd{xiangyu.xie@psi.ch}

\abstract{
  To enhance the spatial resolution of the MÖNCH 25 \textmu m pitch hybrid pixel detector, deep learning models have been trained using both simulation and measurement data.
  Challenges arise when comparing simulation-based deep learning models to measurement-based models for electrons, as the spatial resolution achieved through simulations is notably inferior to that from measurements.
  Discrepancies are also observed when directly comparing X-ray simulations with measurements, particularly in the spectral output of single pixels. 
  These observations collectively suggest that current simulations require optimization.

  To address this, the dynamics of charge carriers within the silicon sensor have been studied using Monte Carlo simulations, aiming to refine the charge transport modeling.
  The simulation encompasses the initial generation of the charge cloud, charge cloud drift, charge diffusion and repulsion, and electronic noise.
  The simulation results were validated with measurements from the MÖNCH detector for X-rays, and the agreement between measurements and simulations was significantly improved by accounting for the charge repulsion.
}


\proceeding{25$^{\text{th}}$ International Workshops on Radiation Imaging Detectors\\
  01-04 July 2024\\
  Lisbon}

\begin{document}
\maketitle
\flushbottom

\section{Introduction}
The MÖNCH detector, currently under development at the Paul Scherrer Institute, is a charge-integrating hybrid pixel detector \cite{MÖNCH}.
Its 25 \textmu m pitch, combined with high dynamic range, fast readout, and low noise, makes the MÖNCH detector a promising candidate for electron microscopy applications.
Our prior study \cite{MÖNCH_DL} demonstrated that deep learning can effectively reconstruct the incident position of single electrons, thereby enhancing the spatial resolution for 200~keV electrons, in the case of which performance is primarily constrained by multiple scattering effects.
The deep learning models were trained using both simulation data and measurement data.
Simulation data are more straightforward to generate with varied parameters, whereas measurement data acquisition is challenging due to the specialized setup required for the electron microscope. 
A primary challenge with measurement data lies in accurately determining the impact point, compounded by issues such as beam current stability and precise calibration.
Additionally, simulations offer more detailed information, such as the electron track, which is not available in the measurements.

However, the small pitch of the MÖNCH detector not only enables high spatial resolution but also presents challenges in accurately modeling charge transport.
These challenges become apparent when comparing simulation-based deep learning models to those trained on measurements. 
In particular, The spatial resolution achieved using simulations is notably inferior to that achieved with measurements.
To address the origin of this discrepancy, we compared measurements and simulations generated by Allpix Squared \cite{Allpix2} for 200 keV electrons and X-rays photons.
Simulations were conducted using the \emph{ProjectionPropagation} module with employing the MÖNCH detector geometry and measurement conditions: a sensor thickness of 320 \textmu m with a 25 \textmu m pitch, a depletion of 30 V and a bias voltage of 90 V for collecting holes, respectively, and a temperature of 293 K.

\begin{figure}
  \centering
  \label{fig:pixel_energy_spectrum_allpix2}
  \subfigure[]{
    \includegraphics[width=0.45\textwidth]{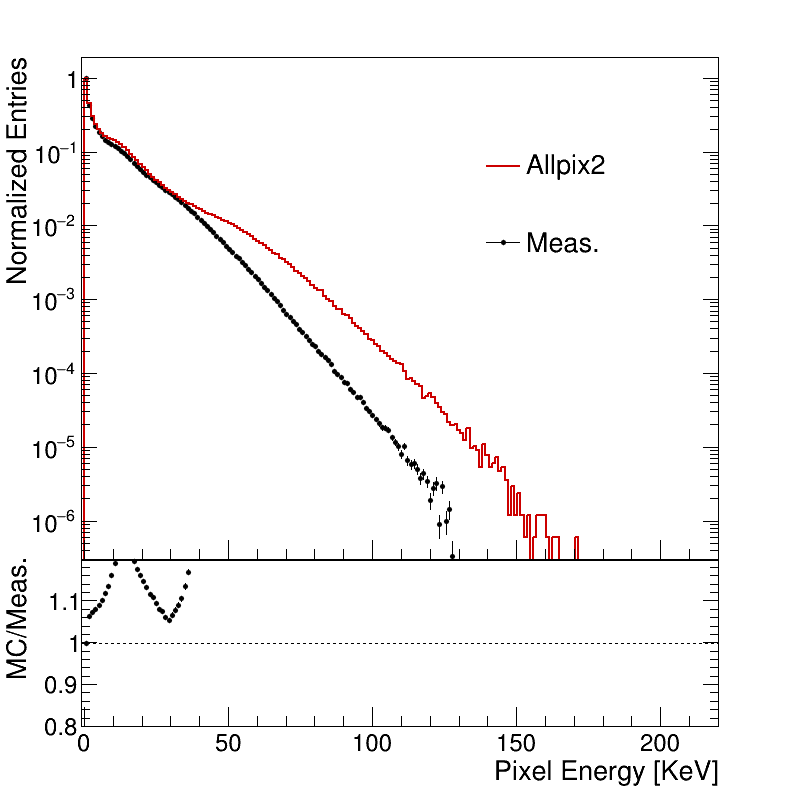}
  }
 
  \subfigure[]{
    \includegraphics[width=0.45\textwidth]{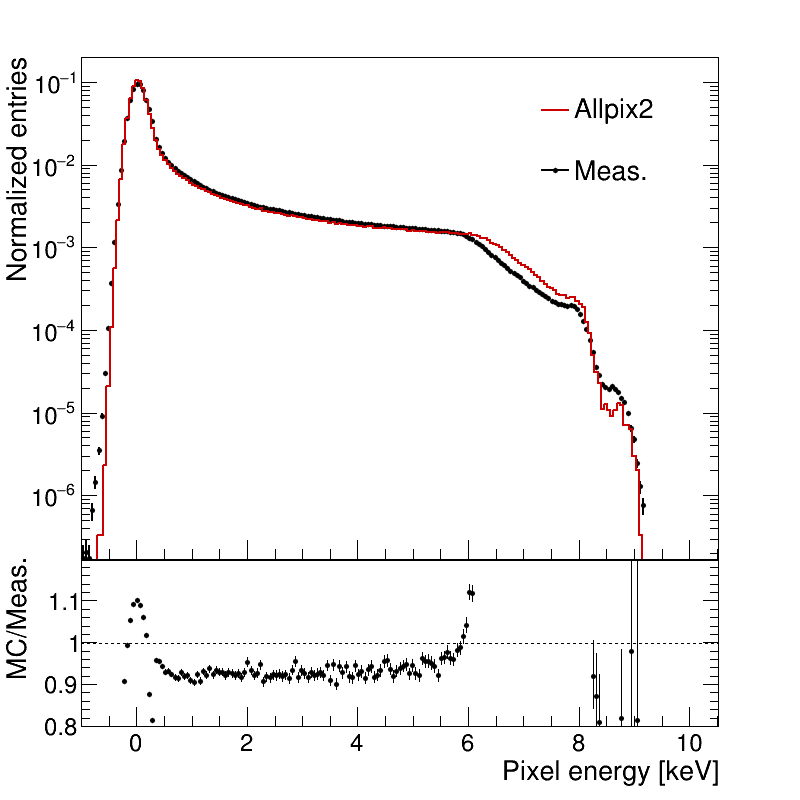}
  }
  \subfigure[]{
    \includegraphics[width=0.45\textwidth]{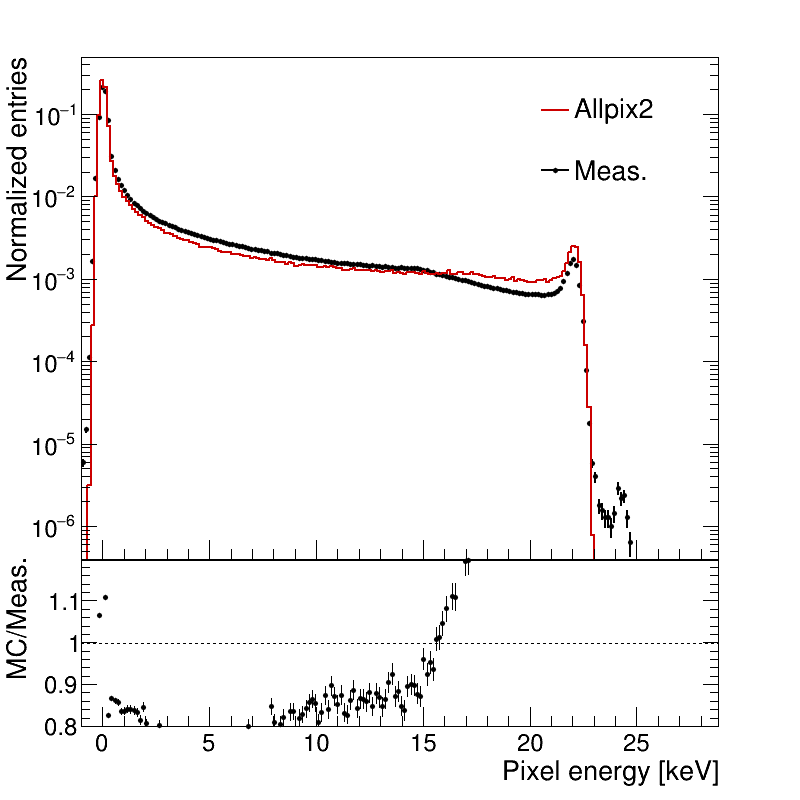}
  }
  \caption{Pixel energy spectra of the MÖNCH detector for 200 keV electrons (a), copper (b), and silver X-ray fluorescence (c) from measurements (black dots) and Allpix Squared simulations (red line). 
  $k\alpha$ X-ray energies are 8.05 keV for copper and 22.16 keV for silver.
  Pixels are those with energy over $5\sigma_{\rm noise}$ for electrons and from 3$\times$3-pixel clusters with a cluster energy within $\pm 0.5$ keV of the K$\alpha$ peak for X-ray photons. 
  The ratio plots between the simulations and the measurements are shown in the lower panels. 
  For consistency across the figures, the ratio range is set to 0.8-1.2.}
\end{figure}

Figure \ref{fig:pixel_energy_spectrum_allpix2} (a), (b), and (c) show the single pixel energy spectra for 200 keV electrons and copper and silver fluorescence X-rays ($\rm E_{k\alpha}=8.05$ keV and $22.16$ keV), respectively.
The selected pixels are those with energy over $5\sigma_{\rm noise}$ for electrons and from 3$\times$3-pixel clusters for X-ray photons.
At higher energies, the simulation shows a larger number of entries than the measurement, as illustrated in the spectra and amplified in the ratio plots. 
This discrepancy suggests an underestimation of the charge-sharing effect in the simulation, which requires optimization.
Furthermore, when comparing the pixel energy spectra for copper and silver X-ray fluorescence (Figure \ref*{fig:pixel_energy_spectrum_allpix2}(b) and (c)), the discrepancies are more pronounced for higher energy depositions.

This study aims to optimize the charge transport simulation for the highly pixelated MÖNCH detector.
A time-stepping Monte Carlo simulation is conducted following the charge carrier dynamics.
The simulation process includes the initial generation of the charge cloud by X-rays or electrons, charge cloud drift, charge diffusion and repulsion, and electronic noise.
The charge repulsion, which is currently not considered in established simulation frameworks \cite{Allpix2,G4Mpx,kDetSim}, is proved to be crucial for accurately reproducing the MÖNCH detector measurements.
The following sections elaborate on the charge carrier dynamics, simulation algorithm, validation results, discussion, and future outlook.

\subsection*{Charge carrier dynamics}
The dynamics of charge carriers in silicon sensors are governed by the continuity equation:
\begin{equation}
  \label{eq:continuity}
    \frac{\partial \rho}{\partial t} = D\Delta\rho - \nabla \cdot (\mu \rho \vec{E})
\end{equation}
where $\rho$ represents the charge carrier density, $D=\mu kT/q$ is the diffusion coefficient, $\mu$ denotes the charge carrier mobility, $k$ is the Boltzmann constant, $T$ is the temperature, $q$ denotes the unit charge, and $\vec{E}$ is the electric field.
The generation and recombination of charge carriers are ignored in the depleted sensor volume.
Driven by the bias voltage, the charge cloud, consisting of charge carriers, drifts towards the readout electrode.
By neglecting charge repulsion, Equation \ref{eq:continuity} simplifies to $\frac{\partial \rho}{\partial t} = D\Delta\rho$, with the coordinate system's origin at the charge cloud center at $t=0$.
The solution to this equation is a Gaussian distribution with a standard deviation of $\sqrt{2Dt}$ in one dimension, a charge diffusion formula widely implemented in simulation frameworks.

However, neglecting charge repulsion in the simulation leads to an underestimation of the charge cloud size, accounting for the discrepancies with measurements in Figure \ref{fig:pixel_energy_spectrum_allpix2}.
To improve the accuracy in modeling charge transport in the MÖNCH detector, charge repulsion must be included in the simulation.
Given that mobility is a complex function of the electric field, it is impractical to find analytical solutions to the continuity equation that account for charge repulsion.
Therefore, this study implemented a time-stepping Monte Carlo simulation to model charge carrier transport in the silicon sensor while incorporating charge repulsion.
X-ray photons are studied due to their point-like energy deposition, which provides an ideal initial condition for simulations.
Additionally, the ease of in-air measurements with metal X-ray fluorescence facilitates a direct comparison between simulations and measurements.

\section{Time-stepping Monte Carlo simulation}
\subsection*{Boundary conditions}
The absorption depth of X-rays in the silicon sensor follows an exponential distribution, with the attenuation length determined by the X-ray energy and the sensor material \cite{attenuationLength}.
The absorbed energy is converted into charge at an efficiency of 3.6 eV per electron-hole pair.

The initial photonelectron follows an semiempirical distribution with a Bethe range \cite{BetheRange_Grun,BetheRange_EH} given by $R_{e^-} = \frac{0.0040}{\rho} \cdot E_{\rm{deposit}}^{1.75} $ \textmu m, where $\rho=2.329$ g/cm$^3$ is the silicon density and $E_{\rm{deposit}}$ is the energy deposition in keV.
This expression is valid within the energy deposition range of 5 to 25 keV.
The 1D standard deviation of the corresponding Gaussian profile is given by $\sigma=\frac{R_{e^-}}{\sqrt{15}}=0.0044\cdot {E_{\rm{deposit}}}^{1.75}$ \textmu m \cite{BetheRangeToSigma}.
The transport begins at the X-ray absorption depth and ends when the charge cloud center reaches the sensor surface driven by the bias voltage.

\subsection*{Charge carrier mobility}
The MÖNCH detector operates in hole-collection mode.
The widely-used Jacoboni-Canali model, which takes into account the saturation of the mobility for high electric fields, is employed to describe the mobility of holes in silicon as a function of the electric field and temperature \cite{Jacoboni-Canali}:
\begin{equation}
  \label{eq:mobility}
  \mu(E)=\frac{v_{m,h}}{E_{c,h}}\frac{1}{(1+(E/E_c)^{\beta_h})^{1/\beta_h}}
\end{equation}
with phenomenological parameters for holes:
\begin{equation}
  \label{eq:mobility_parameters}
  \begin{split}
    v_{m,h} &= 1.62\times 10^8 \ \rm{cm/s}\cdot T^{-0.52}\\
    E_{c,h} &= 1.24\ \rm{V/cm}\cdot T^{1.68} \\
    \beta_h &= 0.46\cdot T^{-0.57}
  \end{split}
\end{equation}
In this study, the temperature is set to 293 K, which is the cooling water temperature for the MÖNCH detector during measurements. The input electric field $E$ is determined by
\begin{equation}
  \label{eq:electric_field}
  E = \sqrt{E_{\rm{drift}}^2+E_{\rm{repulsion}}^2}
\end{equation}
where $E_{\rm{drift}}$ results from the bias voltage and $E_{\rm{repulsion}}$ arises from the charge repulsion, detailed in the following sections.
The calculation assumes orthogonality between $E_{\rm{drift}}$ and $E_{\rm{repulsion}}$, which is valid for charge carriers on the $x$-$y$ plane across the cloud center and is applied to the entire charge cloud.
This assumption maintains spherical symmetry of the charge cloud and simplifies the calculation.

\subsection*{Drift of charge cloud}
In a Cartesian coordinate system with the origin at the charge cloud center and the z-axis aligned with the drift direction, the electric field due to the bias voltage is expressed as:
\begin{equation}
  \label{eq:electric_field}
  E_{\rm{drift}}(z) = \frac{V_{\rm{bias}}-V_{\rm{dep}}}{H}+\frac{2V_{\rm{dep}}}{H}\cdot\frac{z}{H}
\end{equation}
where $V_{\rm{bias}}$ is the bias voltage, $V_{\rm{dep}}$ is the depletion voltage, and $H$ is the sensor thickness.
The depletion voltage of the MÖNCH sensor is measured to be 30 V, and the sensor thickness is 320 \textmu m.
The applied bias voltage $V_{\rm{bias}}$ during measurements is 90 V.

The $z$ coordinate of the charge cloud center is updated as follows:
\begin{equation}
  \label{eq:drift}
  z(t+\delta t) = z(t) + \mu E_{\rm{drift}}(z(t))  \delta t
\end{equation}
where $\mu$ is the average hole mobility of charge carriers and $\delta t$ is the time step, set to 0.01 ns.

\subsection*{Diffusion and repulsion of charge carriers}
The diffusion and repulsion of charge carriers are considered in a Cartesian coordinate system with the origin at the center of the charge cloud. 
For a charge carrier at $(x,y,z)$, diffusion in the $x$ direction is modeled as a random walk process:
\begin{equation}
  \label{eq:diffusion}
  \begin{split}
    x(t+\delta t) &= x(t) \pm \sqrt{2D\delta t} \\
  \end{split}
\end{equation}
where $D=\mu {kT/q}$ is the diffusion coefficient, and the $\pm$ sign indicates that the direction of diffusion is random.
Similar diffusion processes are applied to the $y$ and $z$ directions.

The electric field due to the charge repulsion is calculated as:
\begin{equation}
  \label{eq:repulsion_field}
  E_{\rm{repulsion}}(r) = \frac{Q(r)}{4\pi\epsilon_0\epsilon_r r^2},\ r = \sqrt{x^2+y^2+z^2}
\end{equation}
where $Q(r)$ is the charge within the sphere of radius $r$, $\epsilon_0$ is the vacuum permittivity, and $\epsilon_r$ is the relative permittivity of silicon.
The assumption of spherical symmetry of the charge cloud simplifies the calculation of the electric field due to the charge repulsion.
The $x$ coordinate update due to the repulsion, similar in the $y$ and $z$ directions, is given by:
\begin{equation}
  \label{eq:repulsion}
  \begin{split}
    x(t+\delta t) &= x(t) + \mu E_{\rm{repulsion}}(r) \delta t \cdot x/r \\
  \end{split}
\end{equation}

\subsection*{Intermediate results and parameterization}

Intermediate simulation results from 8.05 keV X-rays (i.e., Cu K$\alpha$ X-fluorescence energy) with an absorption position of (0, 0, 2.5 \textmu m) are presented below.
The Monte Carlo simulation is repeated multiple times to achieve a statistical power of approximately 100,000 charge carriers.
Figure 1(a) plots the root mean square (RMS) of charge carrier $x$ coordinates over drift time with and without the charge repulsion.
The 1D RMS width at the end of the simulation, including the charge repulsion, is 8.46 \textmu m, which is significantly higher than the 1D RMS without the repulsion (7.70 \textmu m).

Parameterization of the charge distribution is crucial for efficiently generating simulation samples.
The final charge distribution, with the repulsion considered, is fitted by a generalized Gaussian distribution:

\begin{equation}
  \label{eq:ggd}
  f(x) = \frac{\beta}{2\alpha\Gamma(1/\beta)}\exp(-\left| \frac{x}{\alpha} \right|^\beta)
\end{equation}
where $\alpha$ is the scale parameter, $\beta$ is the shape parameter, and $\Gamma$ is the gamma function.
The fitted generalized Gaussian distribution is shown as the black curve in Figure 1(b).
The shape parameter $\beta>2$ describes the broadening of the charge distribution due to charge repulsion.
The $\chi^2/\rm{ndf}$ of the fit is 1.0, indicating that the generalized Gaussian distribution effectively models the charge distribution.
Good fits are also observed for different absorption depths and X-ray energies.

\begin{figure}[htb]
  \subfigure[]{
    \includegraphics[width=0.45\textwidth]{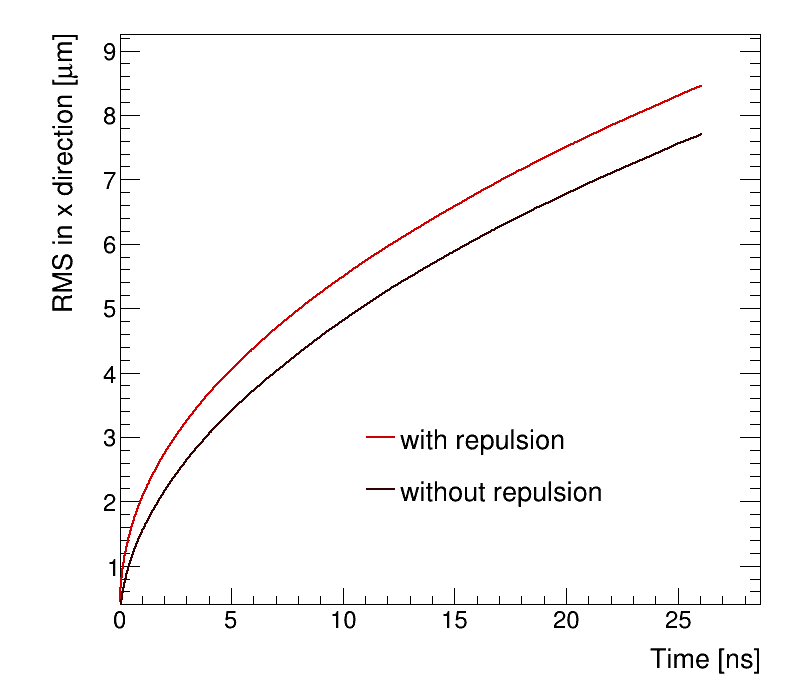}
    \label{fig:rms_over_time}
  }
  \subfigure[]{
    \includegraphics[width=0.45\textwidth]{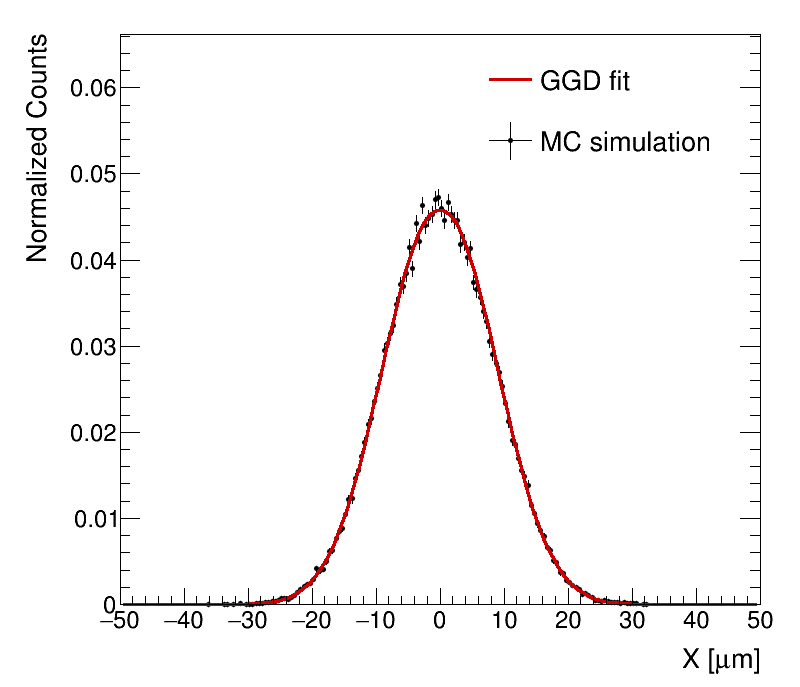}
  }
  \caption{Simulation results of 8.05 keV X-ray photons (Cu K$\alpha$) with an absorption depth of 2.5 \textmu m from the entrance window. (a) The root mean square (RMS) in $x$ over drift time. (b) The charge distribution along the $x$-axis at the end of the simulation with the fitted Generalized Gaussian Distribution (GGD) in red curve.}
\end{figure}

\section{Validation of the simulation}

The validation of the simulation is conducted by comparing the simulation results with measurements obtained using a MÖNCH detector for copper and silver X-ray fluorescence, specifically in terms of the pixel energy spectrum and the charge-weighted center $\eta$ of 3$\times$3-pixel clusters.
X-ray clusters of both the measurements and the simulation follow the same selection criteria and processing steps.
\subsection*{Forming 3$\times$3-pixel clusters}

The previously described simulation is repeated for various absorption depths, with steps of 1 \textmu m.
Then, one million X-rays are simulated, with the absorption depth sampled from an exponential distribution corresponding to the respective attenuation length and a random incident position uniformly distributed in the $x$-$y$ plane.
For each X-ray photon, the charge carrier distribution relative to the cloud center is sampled from the fitted generalized Gaussian distribution for the nearest simulated absorption depth.
Pixel noise, including electronic noise (approximately $0.13\ \rm{keV}$ as determined from measurements) and Poisson fluctuations of the charge carriers, is added to each pixel.
The 3$\times$3-pixel clusters are centered on the pixel with the maximum charge.
Additionally, K$\beta$ X-rays are simulated following the same procedure, with the intensity ratios to K$\alpha$ X-rays given by \cite{kBetaRatio}.

Measurements were conducted using the MÖNCH detector with an X-ray tube featuring copper and silver targets.
To process the raw data from the charge-integration MÖNCH detector, dark current subtraction was applied.
Pixel-wise gain calibration was performed to convert the pixel readout from Arbitrary Digital Units to energy.
The 3$\times$3-pixel clusters were centered on the pixel with the maximum energy.
Pile-up clusters were excluded from the analysis by ensuring that no neighboring pixels of the cluster belonged to another cluster.

A cluster energy selection of $\pm 0.5$ keV of the K$\alpha$ peak was applied to both data sets to further exclude noise and pile-up clusters.

\subsection*{Simulated and measured pixel energy spectra}
\begin{figure}[htb]
  \label{fig:pixel_energy_spectrum}
  \centering
  \subfigure[]{
    \includegraphics[width=0.45\textwidth]{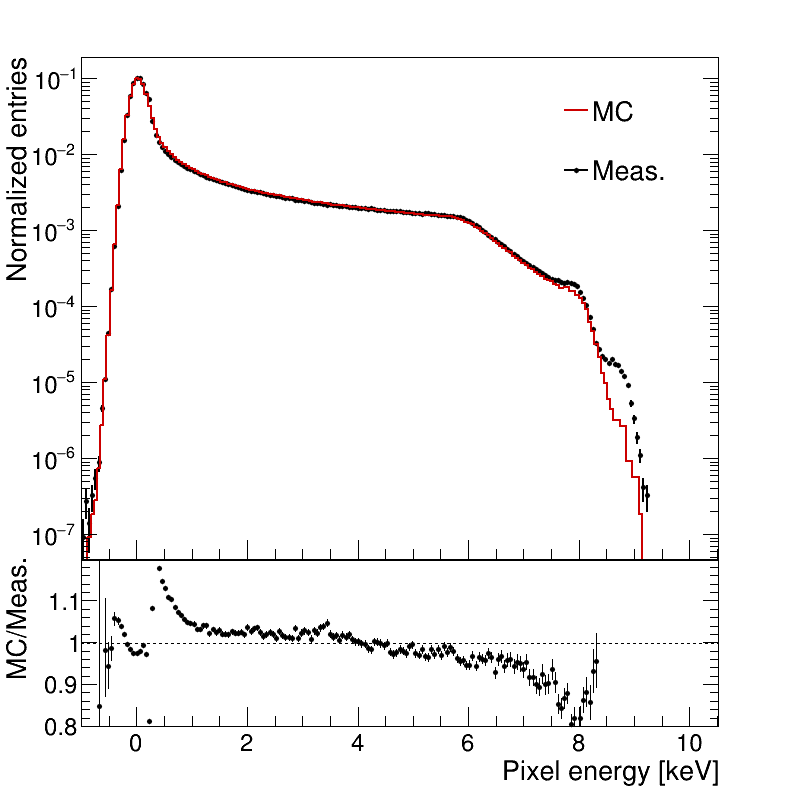}
  }
  \subfigure[]{
    \includegraphics[width=0.45\textwidth]{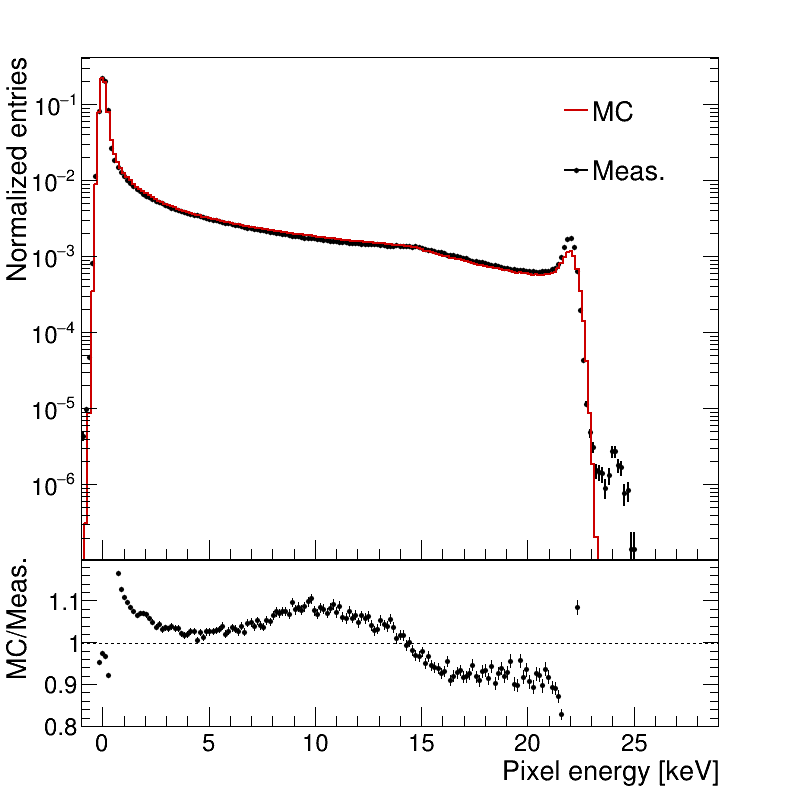}
  }
  \caption{Pixel energy spectrum of the MÖNCH detector for copper X-rays (a) and silver X-rays (b) from measurements (black dots) and simulations (red line). The ratio plots between the simulation and the measurements are shown in the lower panel.}
\end{figure}
The pixel energy spectra of the MÖNCH detector for selected copper and silver X-ray clusters are illustrated in Figure \ref{fig:pixel_energy_spectrum} (a) and (b), respectively.
Both figures demonstrate good agreement between the simulation and the measurements, though some discrepancies remain.
The simulation results show a higher number of entries in the first half of the energy range, while the measurements exhibit more entries in the second half.
These differences are primarily attributed to the simplified noise modeling, as well as non-linearity in charge amplification and pixel cross-talk, which are not accounted for in the simulation.
Overall, significant improvements are observed in the simulation results comparing results obtained without (Figure \ref{fig:pixel_energy_spectrum_allpix2}), and with (Figure \ref{fig:pixel_energy_spectrum}) charge repulsion.

\subsection*{Simulated and measured charge weighted center $\eta_x$, $\eta_y$ parameter distributions}
\begin{figure}[htb]
  \centering
  \subfigure[]{
    \includegraphics[width=0.45\textwidth]{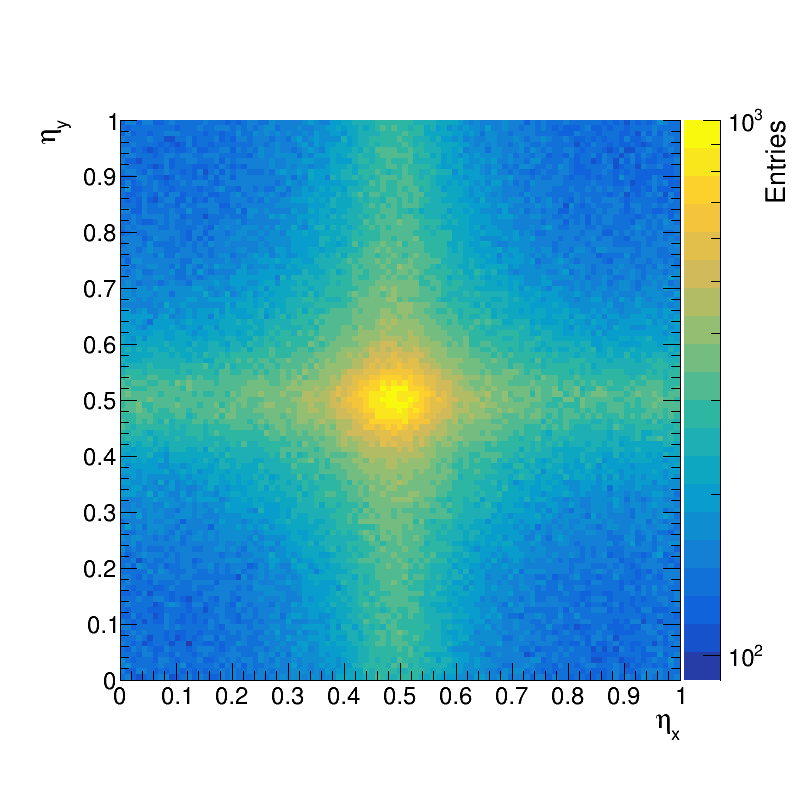}
    }
  \subfigure[]{
    \includegraphics[width=0.45\textwidth]{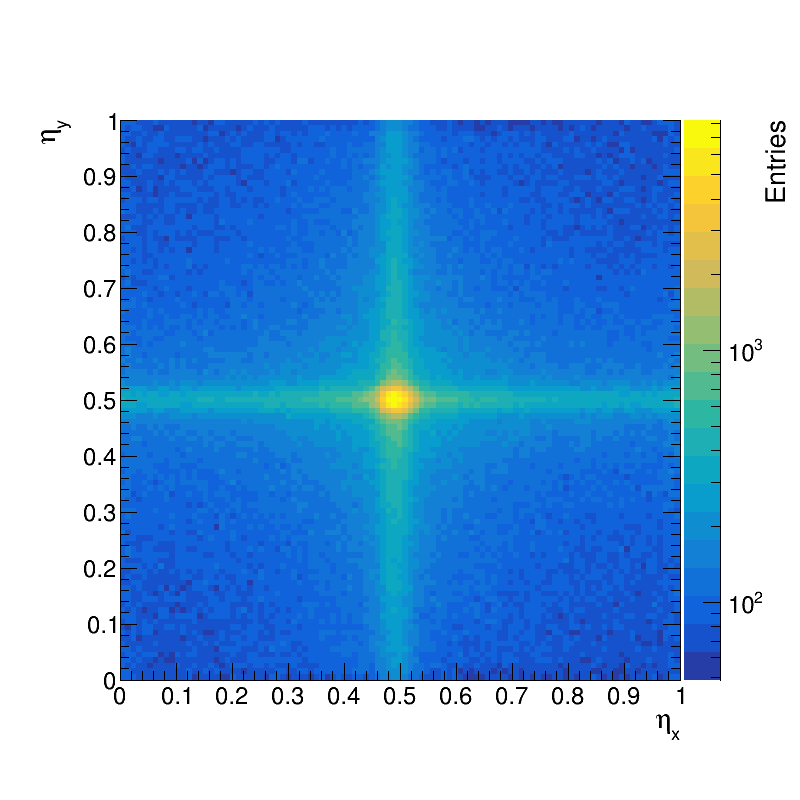}
    }
  \subfigure[]{
    \includegraphics[width=0.45\textwidth]{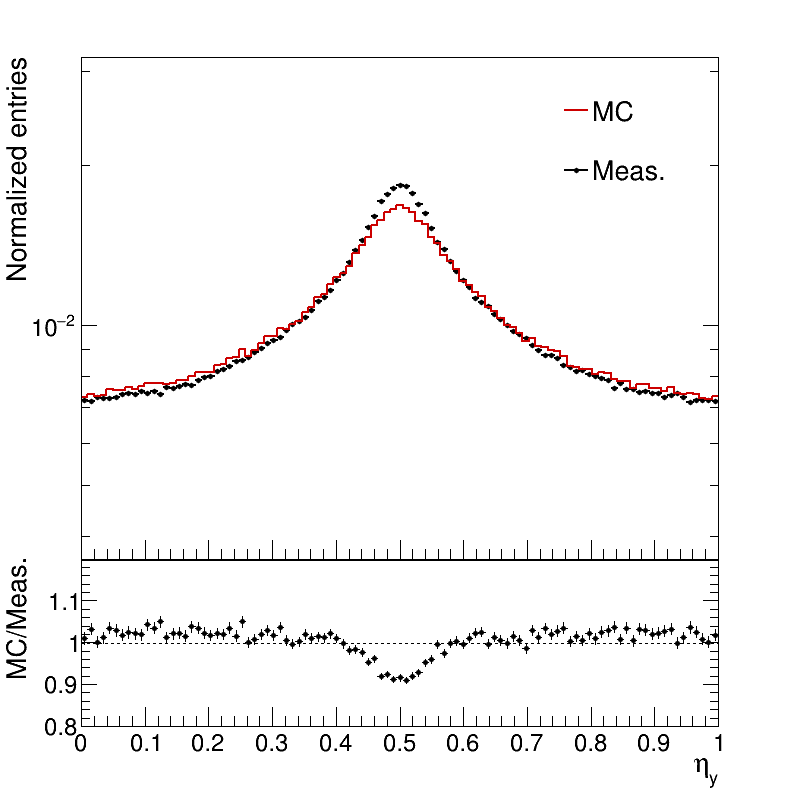}
    }
  \subfigure[]{
    \includegraphics[width=0.45\textwidth]{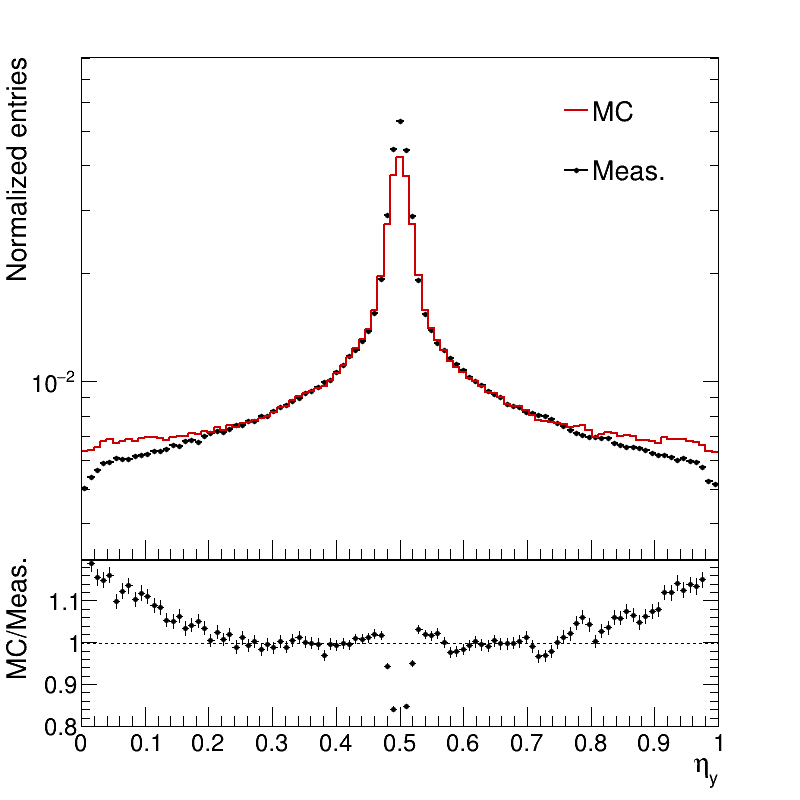}
    }
  \caption{2D histograms of the charge weighted centers $\eta_x$ and $\eta_y$ for copper (a) and silver X-rays (b) from measurements. Projected $\eta_y$ spectra for copper (c) and silver X-rays (d) from measurements (black dots) and simulations (red line). The ratio plots between the simulation and the measurements are shown in the lower panels.}
  \label{fig_charge_weighted_center}
\end{figure}

The charge weighted centers $\eta_x$ and $\eta_y$ are widely used in position interpolation to improve spatial resolution \cite{eta, etaMÖNCH}. 
For the 3$\times$3-pixel clusters, the charge weighted centers $\eta_x$ and $\eta_y$ are calculated as follows:
\begin{equation}
  \label{eq:charge_weighted_center}
  \eta_x = \frac{\sum_{x=0}^{2}\sum_{y=0}^{2}E_{xy}\cdot (x-0.5)}{\sum_{x=0}^{2}\sum_{y=0}^{2}E_{xy}}
\end{equation}
\begin{equation}
  \label{eq:charge_weighted_center}
  \eta_y = \frac{\sum_{x=0}^{2}\sum_{y=0}^{2}E_{xy}\cdot (y-0.5)}{\sum_{x=0}^{2}\sum_{y=0}^{2}E_{xy}}
\end{equation}
Here $E_{xy}$ denotes the energy measured by the pixel\footnote[1]{Charge and energy are translated using a conversion coefficient of 3.6 eV per electron-hole pair.} at the $x$-th row and the $y$-th column.
With the origin of the coordinate system defined as the lower left corner of the center pixel, the charge weighted centers $\eta_x$ and $\eta_y$ range from 0 to 1.

The 2D histograms of the charge weighted centers $\eta_x$ and $\eta_y$ for copper and silver X-rays are depicted in Figures \ref{fig_charge_weighted_center}(a) and (b,) respectively. 
The color bar in logarithmic scale represents the number of entries.
The $\eta_y$ spectra after projection, obtained from both measurements and simulations, are displayed in Figure \ref{fig_charge_weighted_center}(c) and (d).
The simulation aligns well with the measurements for copper X-rays, while there remain some discrepancies for silver X-rays.
The possible sources of these discrepancies are detailed in the following section.

\section{Discussion, conclusion and outlook}

The 25 \textmu m pitch of the MÖNCH detector presents the challenge of a more accurate modeling of charge transport, while simultaneously offering a unique platform for the investigation of this phenomenon.
Meanwhile, we should note that the difference in the charge distribution with and without repulsion, as exampled in Figure 1(a), is less noticeable for hybrid pixel detectors with a larger pixel pitch, such as 55 \textmu m of the Medipix detectors \cite{Medipix} and 75 \textmu m of the Jungfrau detectors \cite{Jungfrau}.

\subsection*{Sources of remaining discrepancies}
While incorporating the charge repulsion into the simulation has significantly improved the agreement between the simulation and the measurements, some discrepancies persist.
The simulation assumes a spherically symmetric charge distribution to simplify the calculation of the electric field due to charge repulsion, an assumption which does not strictly hold.
Additionally, the gradient of the drift electric field within the charge cloud is neglected in the simulation.
When the charge cloud approaches the readout electrode, the symmetry of the charge cloud is broken.
Furthermore, mobility uncertainty is up to 10\% in the Jacoboni-Canali model \cite{Jacoboni-Canali}, which may contribute to the discrepancies.

When comparing the results of the copper and silver X-rays, the discrepancies are more pronounced for the silver X-rays.
Note that the impact of charge repulsion mainly occurs within the first few nanoseconds and then saturates (see Figure \ref{fig:rms_over_time}).
The saturation time also influences the charge distribution.
For silver X-rays, more photons are absorbed in the deeper sensor volume, which amplifies the effect mentioned above.
One possible reason for the differing saturation times is the absence of electrons, the other type of charge carrier, in the simulation, which accelerates the development of the repulsion effect.

In the measurements, uncertainties from measured parameters, such as the depletion voltage, contribute to these discrepancies.
Non-linearity of charge amplification and cross-talk between pixels should also be considered, as discussed previously.

\subsection*{Conclusion and outlook}
In this study, a time-stepping Monte Carlo simulation incorporating charge repulsion was developed to simulate charge carrier transport in the silicon sensor.
Parameterization was implemented to effectively model the charge distribution and efficiently generate simulation samples.
The simulation results were validated against measurements obtained from the MÖNCH detector for X-rays.
By considering charge repulsion, the consistency between the simulation and the measurements has been significantly improved.

Further research is necessary to implement this approach for electron simulations. 
Modeling the charge cloud for electrons is more challenging than for X-rays due to the continuous energy deposition spectrum.
The parameters of the charge cloud distribution should be modeled as a function of both the deposition depth and the deposition energy.
Additionally, the repulsion between charge clouds must be considered.
With optimized simulations, higher-quality simulation samples can be generated for deep learning applications.
Various design parameters of the MÖNCH detector, e.g., the sensor thickness and the bias voltage, can be investigated to exploit the charge sharing effect for a better spatial resolution.

\subsection*{Code availability}
TThe simulation source code, along with an example Jupyter notebook that includes fitting and plotting scripts, as well as histograms of the measurement results, can be accessed at \url{https://github.com/slsdetectorgroup/ChargeTransportSimulation}.

\subsection*{Acknowledgements}
The authors would like to express their gratitude to Dr. Emiliya Poghosyan and Dr. Elisabeth Müller for their expertise in conducting measurements with the electron microscope.

\end{document}